\documentclass[epj]{svjour}
\usepackage{graphicx}
\usepackage{epstopdf}
\usepackage{times,amsmath}
\usepackage{helvet}
 \newcommand{\D}{\textstyle{\rm d}}
 \newcommand{\E}{\textstyle{\rm e}}

\input epsf
\begin{document}

\hugehead
\title{\Large Source radii at target rapidity from two-proton and two-deuteron correlations in central Pb+Pb collisions at 158 {\it A} GeV}
 \subtitle{WA98 Collaboration}
\author{M.M.~Aggarwal\inst{1} \and Z.~Ahammed\inst{2}\and A.L.S.~Angelis\inst{3,*}
\and V.~Antonenko\inst{4}\and V.~Arefiev\inst{5}\and
V.~Astakhov\inst{5}\and V.~Avdeitchikov\inst{5}\and
T.C.~Awes\inst{6}\and P.V.K.S.~Baba\inst{7}\and
S.K.~Badyal\inst{7}
% ...
}

\institute{University of Panjab, Chandigarh 160014, India \and
Variable Energy Cyclotron Centre,  Calcutta 700 064, India \and
University of Geneva, CH-1211 Geneva 4,Switzerland \and RRC
Kurchatov Institute, RU-123182 Moscow, Russia \and Joint Institute
for Nuclear Research, RU-141980 Dubna, Russia \and Oak Ridge
National Laboratory, Oak Ridge, Tennessee
  37831-6372, USA \and
University of Jammu, Jammu 180001, India
% ...
}

\date{Received: date / Revised version: date}
%

 \abstract{
Two-proton and
two-deuteron correlations have been studied in the target fragmentation
region of  central Pb+Pb collisions at 158 $A$ GeV. Protons
and deuterons were measured with the Plastic Ball
spectrometer of the WA98 experiment at the CERN SPS. 
The results of one-dimensional and
multi-dimensional analyses using both the Bertsch-Pratt and
Yano-Koonin-Podgoretsky parameterizations of the two-particle
correlation functions are presented. The proton source exhibits a volume
emission, while the deuteron source, with small outward
radius, appears opaque. Both proton and deuteron
sources have cross-terms $R_{ol}^2$ and longitudinal velocities
$\beta$ consistent with zero, indicating a boost-invariant
expansion. The invariant radius parameter $R$ follows an approximate 
$A/\sqrt{m}$ scaling while the longitudinal and transverse radii,
 $R_{L}$ and $R_{T}$, scale approximately as $A/\sqrt{m_{T}}$
with $A\approx 3$ fm~GeV$^{1/2}$ in both cases.
}
%
\maketitle 
\section{Introduction}
\label{intro}

In this paper we report on an analysis of quantum-statistical and final state induced correlations
\cite{wiedeman} of proton and deuteron pairs emitted from 158 $A$
GeV central Pb+Pb collisions. The study of these two particle species
in the same system is of special interest as one would expect
that deuterons, due to their loose structure and small binding
energy, will survive only in the late, low density
environment when scatterings are rare. The generally 
accepted mechanism of deuteron production is by
coalescence of a proton and neutron 
\cite{csernai,polieri,polieri1}  at the time of freeze-out. The deuteron
measurement thus provides
an unique means to measure the geometry of the
source in the late time interval of the expansion. 
Experimental results on Fermi-Dirac correlations for protons are not as
abundant as those on Bose-Einstein correlations for mesons, and are 
very scarce for deuteron correlations.
A remarkable feature of the existing proton and deuteron correlation data in
hadron-nucleus and nucleus-nucleus collisions, spanning over four
orders of magnitude of  incident energy from 0.02 to 450  $A$
GeV is the small
variation of the extracted source radius parameter~\cite{FOX,GONG,CEBRA,KOTTE,BARTKE1,AGAKISHIYEV,WA80,NA44,NA49,GUSTAFSSON,BOAL,CHEN,POCHODZALLA,POCHODZALLA1986,LYNCH1983,AWES1988,CHITWOOD}.

\section{Experimental setup}
\begin{figure}[ht]
\includegraphics[width=90mm]{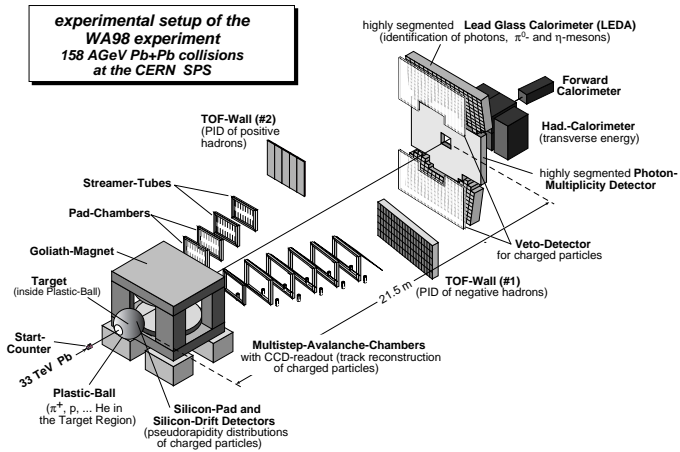}
\caption[x]{\small  Experimental setup
of the WA98 experiment for the 1996 run.}
 \label{wa98-setup}
 \end{figure}

The CERN SPS experiment WA98 (for details see \cite{WA98} and
references therein) is a general purpose apparatus that consists
of large acceptance photon and hadron spectrometers, detectors for
charged particle and photon multiplicity measurements, and
calorimeters for transverse and forward energy measurements.
The layout of the WA98 experimental setup for the
1996 SPS run period is shown in Fig.~\ref{wa98-setup}.
The results presented in this report were obtained from an analysis of
data taken in 1996 with the 158 $A$ GeV $^{208}$Pb beam 
on a 239 mg/cm$^2$ $^{208}$Pb target and  made use of the
Midrapidity Calorimeter(MIRAC)~\cite{MIRAC}, the Zero Degree
Calorimeter(ZDC)~\cite{ZDC}, and the Plastic Ball spectrometer~\cite{PBALL}. 

The ZDC registers energy emitted
along the beam direction in the $3^\circ$ forward cone. The MIRAC
measures the total transverse energy in the pseudorapidity region
of $3.5 \le \eta \le 5.5$. It is a sampling calorimeter that  consists 
of a lead-scintillator electromagnetic
section, followed by an
iron-scintillator hadronic section. MIRAC plays the central role
in the WA98 minimum bias trigger where the measured transverse
energy $E_T$ is required to be above a minimum threshold.  MIRAC
is used to classify the centrality of each event. The analysis 
presented here has been performed on the 10\% most central 
events of the measured minimum bias cross 
section of $\approx 6400$ mb.  About 3 millions central Pb+Pb events were
analyzed with an average of 6.1 identified protons and 2.9  
identified deuterons per event.

\begin{figure}[ht]
\includegraphics[width=90mm]{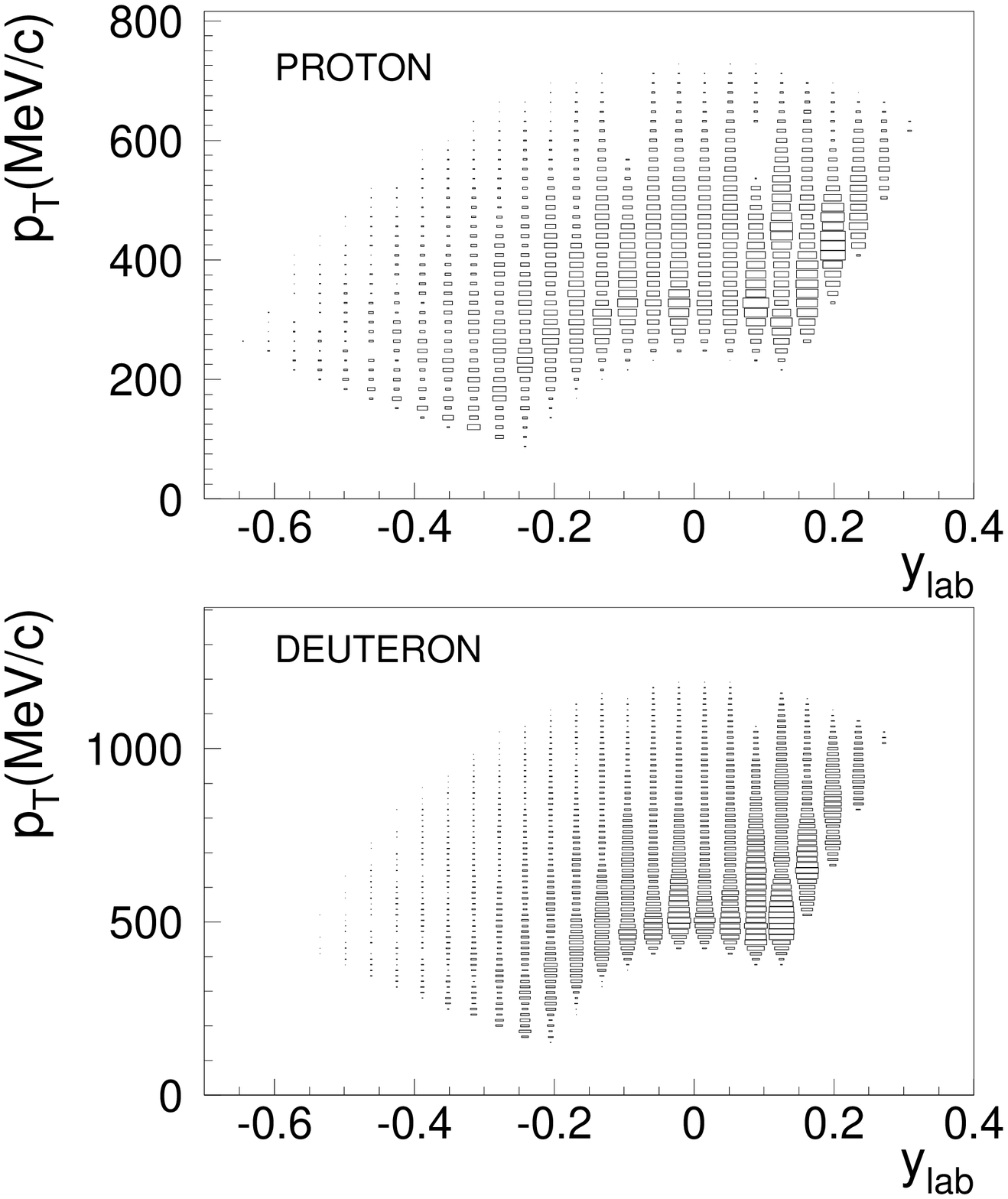}
\vskip -1.2cm
\caption[x]{\small The transverse momentum
versus rapidity distribution for protons and deuteron registered
in Plastic Ball spectrometer.} \label{y-pt}
\end{figure}

The identification and momentum measurement of protons and
deuterons has been made using the Plastic Ball spectrometer~\cite{PBALL}.
The Plastic Ball consists of 655 
detector modules surrounding the target over the range of
polar angles $\vartheta$ from $30^{\circ}$ to $160^{\circ}$
with full azimuthal acceptance. 
Each of the modules consists of a $\Delta E$ and $E$ section. The
$\Delta E$ section is a CaF$_2$(Eu) crystal scintillator and the $E$
section is a plastic scintillator. The light emission of the
plastic scintillator is approximately 100 times faster than that
of the CaF$_2$(Eu), so signals from both scintillators can be read
out by a single photomultiplier with subsequent separation of the
signals by pulse shape analysis. The $\Delta E-E$ identification
technique was used and is capable of identifying
pions and charged fragments up to the helium isotopes.
Because of the relatively low yield of pions as compared to
baryons, an additional positive pion identification is made by
means of its sequential decay $ \pi^{+} \rightarrow \mu^{+}
(\nu_{\mu}) \rightarrow e^{+} (\nu_e \bar{\nu_{\mu}}) $.
The thickness
of the $\Delta E$ crystal scintillator was chosen to be 4~mm with
35.6~cm as the length of the $E$ plastic scintillator. This assures
that protons up to 200 MeV are fully stopped in the $E$-counter and
provide a complete $\Delta E-E$ signal. Additionally,
the mis-identification of punch through deuterons do not disturb
the proton spectra below 200 MeV. 
 
The WA98 experimental setup has been implemented in the GEANT
Monte Carlo simulation package~\cite{GEANT}. This has been used
to  study the response of the Plastic Ball in order to extract  the particle
identification efficiency and the kinetic energy 
resolution for different particle species, as well as two-particle variables 
and takes into account the granular structure of the Plastic Ball. 
GEANT transports the particles through the experimental setup taking 
into account the geometrical material boundaries and particle 
interactions. The particles are traced through the experimental setup 
where they may interact with the material they encounter on their way 
to the detector. When a particle reaches the detector the output signal 
of the Plastic Ball module is simulated. 
The particle distributions used as input to the GEANT simulations 
were adjusted to reproduce the measured momentum 
and angular distributions of particles. The Plastic Ball acceptance 
is illustrated in Fig.~\ref{y-pt} as the distribution
of protons and deuterons that pass the acceptance cuts.

\section{One-dimensional analysis}

The one-dimensional experimental correlation function was
constructed as:
\vskip -0.3cm
\begin{equation}
 C(q)=N\;\frac{{Y_{12}(q)} }{{Y^{*}_{12}(q)}}
\end{equation}
where ${\mathbf q}=\frac{1}{2}(\mathbf{q_1}-\mathbf{q_2})$ is the two-particle three-momentum 
difference in the laboratory frame, $N$ is a
normalization constant,  and the numerator is the coincidence
yield while the denominator is the uncorrelated background yield.
In this analysis, the event-mixing technique was used to construct
the background yields, using the same central event data set as used to 
construct the numerator. 
To insure that only particles from comparable events were mixed, the central
event sample was further subdivided into 8 subsets according to the 
measured transverse energy. 
Particles were picked randomly from different
events belonging to the same centrality selection 
with the additional condition to
have  the same multiplicities of the
studied particle and all particles registered in the Plastic
Ball. The number of mixed background pairs was
chosen to be 10 times larger than the number of coincident pairs to 
insure a small statistical error contribution from the background pair yield, $Y^{*}_{12}$. 

The measurement resolution of the variable $q$ 
in the Plastic Ball spectrometer had an
average value of $\sigma (q)\simeq 10.5$ MeV/c
and 12.5 MeV/c for the two-proton and two-deuteron systems, respectively. For small
values of $q$, the resolution was around 7.5 MeV/c for the two-proton system. 
A bin size of 7.5 MeV/c was chosen for the one-dimensional correlation
analyses. It should be noted that the one-dimensional correlation
functions presented here have been studied as a function of the momentum difference $q$
rather than the more commonly  used invariant four-momentum difference
$q_{inv}=\frac{1}{2}\sqrt{-(p_1^{\mu}-p_2^{\mu})^2}$. The experimental 
resolution for the pp (dd) system in $q_{inv}$ at 20 (40) MeV/c is about 15 (20) MeV/c. 
It increases to 20 (27) MeV/c at $q_{inv}$=100 (200) MeV/c and approximately levels off
above that.

\subsection{Final state interaction}
To calculate the pair-wise final state interaction we used two computer codes for
the proton-proton system: (i) the CRAB code written by S.Pratt \cite{wwwPratt} which includes
Coulomb and strong final state interactions, and (ii)  the static model developed by A.
Deloff \cite{deloff1,deloff2} with both Coulomb and strong S-wave
interactions included. The latter incorporates a potential of the
delta-shell form
\begin{equation}
    \label{eq:2}
    2\mu V(r)= -(s/R)\,\delta(r-R),
\end{equation}
characterized by the range $R$ and the
dimensionless parameter $s$ representing the strength of the force ($\mu$ is the reduced
mass). Only the $^1S_0$ 
partial wave in the strong interaction was retained. This pp interaction is
well known experimentally and the measured phase shifts can
be satisfactorily reproduced up to $q$=150 MeV/c by taking the delta-shell 
potential parameters as $s$=0.906 and $R$=1.84 fm.
Both codes gave
consistent results for the proton-proton system. 

\begin{figure}[ht]
\vskip -2.0cm
\includegraphics[width=90mm]{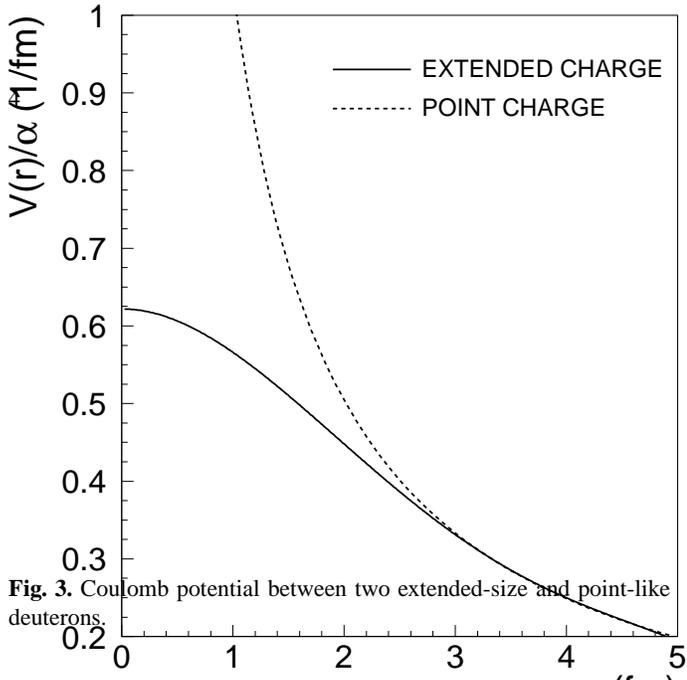} 
\vskip -2.0cm
 \caption[x]{\small Coulomb potential between two extended-size
and point-like deuterons. } \label{poten}
\end{figure}

For the deuteron-deuteron case the model by A. Deloff \cite{deloff1}
was used in which the strong s-wave interaction in S=0,2 spin states
was included together with
the Coulomb repulsion between two extended-size deuterons. 
The Coulomb interaction between
two deuterons is different from that  between two protons because
the deuteron is composite and relatively large. For two extended-size
deuterons located at $\mathbf {r}_1$ and $\mathbf {r}_2$, the
Coulomb potential can be written as
\begin{eqnarray}
V(|{\mathbf r}_1-{\mathbf r}_2|)=\alpha\int \D^3{\mathbf
x}_1\D^3{\mathbf x}_2\, \frac{\rho(|{\mathbf x}_1-{\mathbf
r}_1|)\rho(|{\mathbf x}_2-{\mathbf r}_2|)}{|{\mathbf x}_1-{\mathbf
x}_2|} \label{potent}
\end{eqnarray}
where $\rho(x)$ is the charge density and $\alpha$ is the fine
structure constant. Since the right hand side of Eq.~(\ref{potent}) is
translationally invariant, the potential $V(r)$ depends only
upon the difference $r=|\mathbf{r}_1-\mathbf{r}_2|$, and for $r$
larger than twice the deuteron radius takes the usual point-like
form $V(r)=\alpha/r$. For an assumed uniform charge distribution, 
this integral can be obtained in analytic form as:
\begin{equation}
\label{potential} V(r)= \left \{
\begin{array}{ll}
\frac{\alpha}{R_c} \, \frac{1}{x} \{ 1-3(1-x)^4  &  \, \\
\,\,\,\,\,\times [1-\frac{2}{15}(1-x)(5+x)]\} \ \ \ & \mbox{$0 \leq x \leq 1$} \\
\frac{\alpha}{r} & \mbox{$x > 1$}
\end{array}
\right.
\end{equation}
where  $x=r/R_C$ and $R_C=3.86$ fm is twice the deuteron radius. A
comparison of the Coulomb potential for extended-size and point-like
deuterons is shown in Fig.~\ref{poten}. The Coulomb repulsion is observed 
to be substantially reduced when the extended deuteron size is taken
into account.

The one-dimensional two-particle correlation calculation was based on a
static spherical gaussian shaped source, specified by radius $R_0$:
\begin{equation}
\label{brak}
\rho(x)=(2\pi\,R_0^2)^{-3/2}\;\E^{\textstyle-x^2/2R_0^2}.
\end{equation}
In the case of a non-zero emission lifetime this approach gives an {\em upper} limit for
the spatial extent of the source.

To account for the Plastic Ball resolution all of the calculated
correlation functions presented in this paper were calculated as:
\begin{equation}
C(q)=\int r(q,q^{\prime}) C(q^{\prime}){\rm d}q^{\prime}
\end{equation}
with the predicted correlation function $C(q^{\prime})$ convoluted with the Gaussian resolution
function of the Plastic Ball $r(q,q^{\prime})$.
\begin{equation}
r(q,q^{\prime})={1\over(2\pi)^{3/2}}\  {1\over \sigma(q)}\  {\rm
e}^{-{1\over 2} {(q-q^{\prime})^2 \over \sigma(q)^2}} {\rm
d}q^{\prime}
\end{equation}
as determined from the GEANT simulations.

\subsection{Systematic errors}
The systematic errors have been evaluated by comparison of the results 
obtained under different conditions. The systematic uncertainties are 
mainly introduced by the granularity of the detector and, to some extent, 
by the procedure used to construct the background pair distribution. The following 
sources of systematic errors have been investigated leading to variations 
(rms estimates) of the radius parameters given in brackets, with the first and second 
numbers referring to the pp and dd system, respectively:

\begin{itemize}
\item[(i)]{ Different periods of data taking during the run [0.05, 0.05 fm].}

\item[(ii)]{ Different particle identification windows  [0.09, 0.12 fm].}

\item[(iii)]{  The width of the transverse energy $E_T$ bin used for centrality
selection in the background pair distribution calculations: the centrality
bin size was halved and the result was compared to the nominal bin size result [0.01, 0.05 fm].}

\item[(iv)]{ Non-conservation of momentum in the background pair correlations:
local (transverse) momentum conservation in real events, although not 
expected to be exact, may influence the correlation between real particle
pairs due to the fact that a significant fraction, but not all, of the particles in
the target rapidity region are measured in the Plastic Ball.  
To investigate the sensitivity of the fitted
parameters to the effect of momentum conservation, in addition to the requirement
that the full background event have the same number of particles
as the real event detected in the Plastic Ball, it was further required that the
background event have a total momentum of all measured particles
equal to that of the real event:
$P_{Tot}({\rm measured})=P_{Tot}({\rm reference})\pm \Delta$, where  $\Delta$ should
be 0 in the ideal case. In practice, $\Delta$  was chosen to be 300 MeV/c
and 600 MeV/c for the two-proton and two-deuteron background
events, respectively [0.07, 0.06 fm].}

\item[(v)]{ The granular structure of the Plastic Ball spectrometer:
the GEANT program has been used to study the distortion of the correlation 
function due to the granularity of the Plastic Ball detector  [0.15, 0.10 fm].}

\item[(vi)]{ The bin width of the correlation function:  to examine the dependence of
the radius parameters on the $q$ bin width, the analysis was also performed
with bin widths of 10 MeV/c and 5 MeV/c [0.06, 0.06 fm].}

\item[(vii)]{ Detector variation: the analysis was performed separately in
two different azimuthal angle intervals: $0^\circ - 180^\circ$ and
$180^\circ - 360^\circ$ [0.01, 0.02 fm].}

\item[(viii)]{ The effect of possible mis-identification of protons and
deuterons due to the multiple hits to the same Plastic Pall module.
Since the multi-hit
probability depends very strongly on $\theta$, the analysis was performed
with two subsamples with $\theta$ intervals of protons and deuterons: $60^\circ -90^\circ$
and $90^\circ-160^\circ$. The ratio of contaminations due to
multi-hits in these $\theta$ intervals is about 10:1. The GEANT
simulations also demonstrated that the contamination from spurious
protons and deuterons contribute uniformly to their correlation
functions. As a result, this background contamination does not
affect the shape of the correlation function, leading to the same,
within errors, radius parameters [0.15, 0.10 fm].}

\end{itemize}

 The total systematic uncertainty was calculated by adding in quadrature the contributions
 listed above.

\subsection{Two-proton system}

\begin{figure}[ht]
\vskip -2.0cm
\includegraphics[width=90mm]{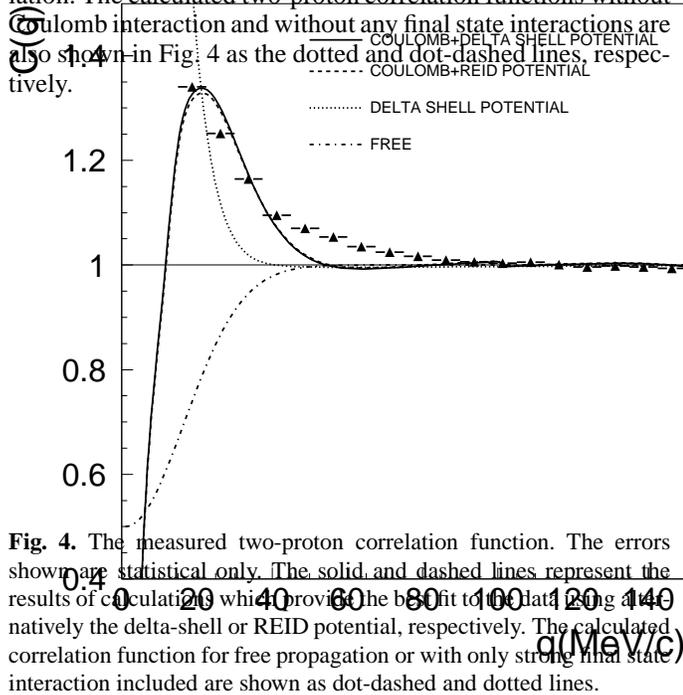} 
\vskip -2.0cm 
\caption[x]{\small
 The
measured two-proton correlation function. The errors shown are
statistical only. The solid and dashed lines represent the results
of calculations which provide the best fit to the data using alternatively
the delta-shell or REID potential, respectively. The calculated correlation
function for free propagation or with only strong final state interaction
included are shown as dot-dashed and dotted lines. }
\label{p-corr}
\end{figure}

The measured two-proton correlation function is shown in 
Fig.~\ref{p-corr}. The solid line represents the result of the
theoretical calculation which gives the best fit to the data 
using the delta-shell potential.  The final state Coulomb and
strong interactions, the Fermi-Dirac effect, and the experimental
resolution have been implemented in the theoretical calculation.
The resulting radius parameter is 3.14$\pm$ 0.03(stat.)$\pm$
0.21(syst.) fm.

Although the
delta-shell potential may not be realistic, its main advantage is that
both the phase shift and the enhancement factor can be obtained in
an analytic form. Furthermore, it is desirable to use the same model for
calculation of both proton and deuteron correlation functions. 
It was demonstrated in \cite{deloff2}
that  the calculated pp phase shifts as
obtained from both the delta-shell potential and the REID soft core
potential \cite{REID} agree well with the experimental data, up to about 
150 MeV/c. As a
cross check, the REID potential was also used to fit the
experimental two-proton correlation function. The resulting correlation function is shown in
Fig.~\ref{p-corr} as the dashed line. The fitted radius parameter
is 3.05 $\pm$ 0.02(stat.)$\pm$ 0.24(syst.) fm which, within errors,
is consistent with the delta-shell potential calculation. The
calculated two-proton correlation functions without 
Coulomb interaction and without any final state interactions
are also shown in Fig.~\ref{p-corr} as the dotted and
dot-dashed lines, respectively.

\begin{figure}[ht]
\vskip -2.0cm
\includegraphics[width=90mm]{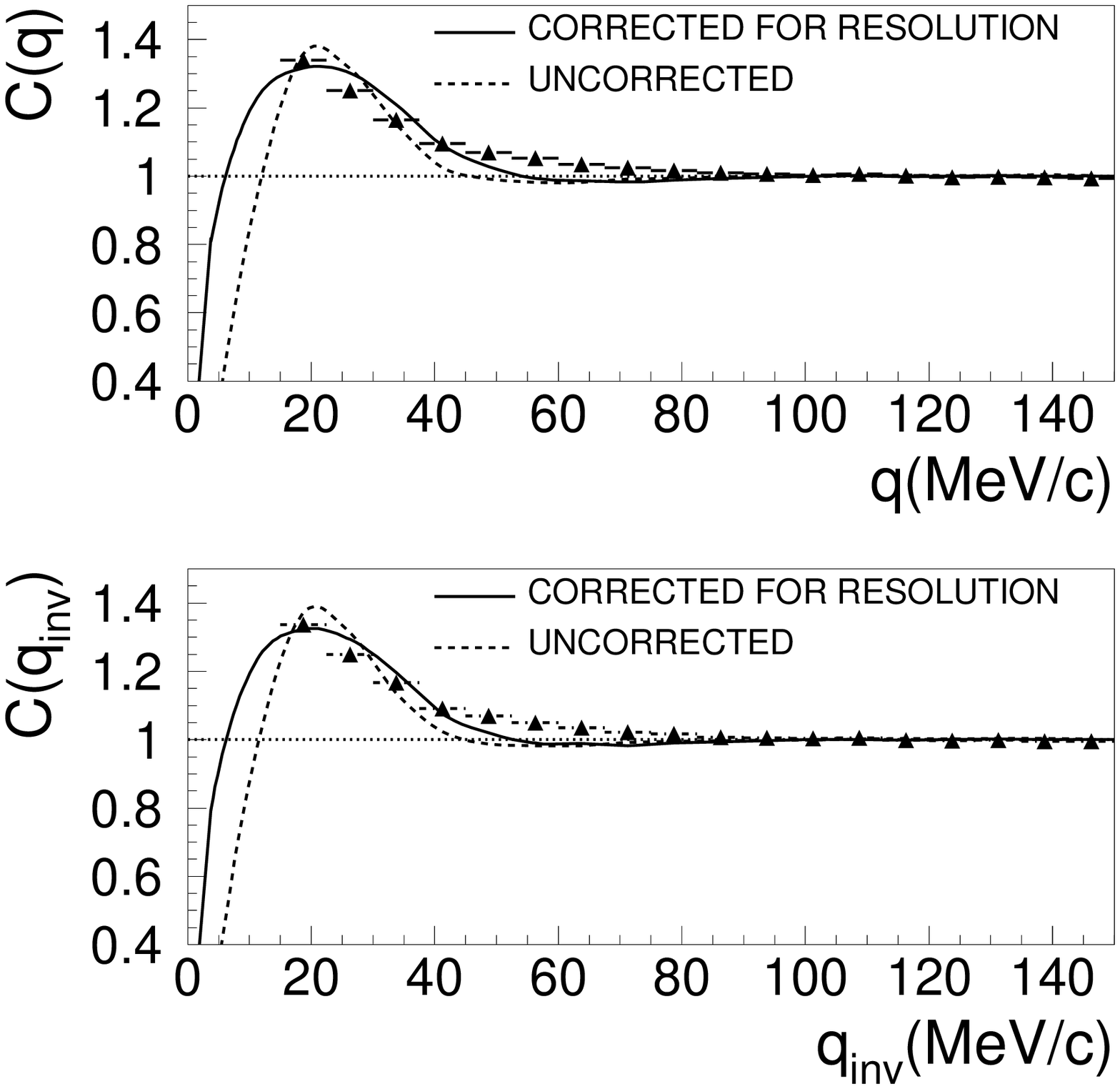} 
\vskip -2.0cm 
\caption[x]{\small The
two-proton correlation function as a function of $q$ and
$q_{inv}$. Only statistical errors are shown.
The solid and dashed lines are results from the CRAB
calculation with and without the experimental resolution applied, 
respectively.  } \label{crab-q-qinv}
\end{figure}

In order to
make a direct comparison with other experimental results, the CRAB
package~\cite{wwwPratt}, provided by Pratt, was also used to extract the radius
parameter for the two-proton system. The CRAB calculation takes
into account the Coulomb and strong final state interactions. To
describe the latter for the two-proton system the REID potential was
taken. The measured momentum distribution of protons was used as
input to the CRAB calculation. To obtain a radius parameter, 
the calculation was performed for a set
of different radius parameters and then compared to the
experimental data by calculating the $\chi^2$ value. This gave a
best fit radius parameter of $2.83\pm 0.20$ fm, that is consistent, within errors, 
with the delta-shell model calculation. The CRAB calculation result is
shown in Fig.~\ref{crab-q-qinv}. The experimental correlation as 
a function of $q_{inv}$  is also shown in Fig.~\ref{crab-q-qinv} with the 
CRAB calculation. 
The solid and dashed lines represent the CRAB calculations with
and without the experimental resolution effect included, respectively.

\begin{figure}[ht]
\vskip -1.0cm
\includegraphics[width=90mm]{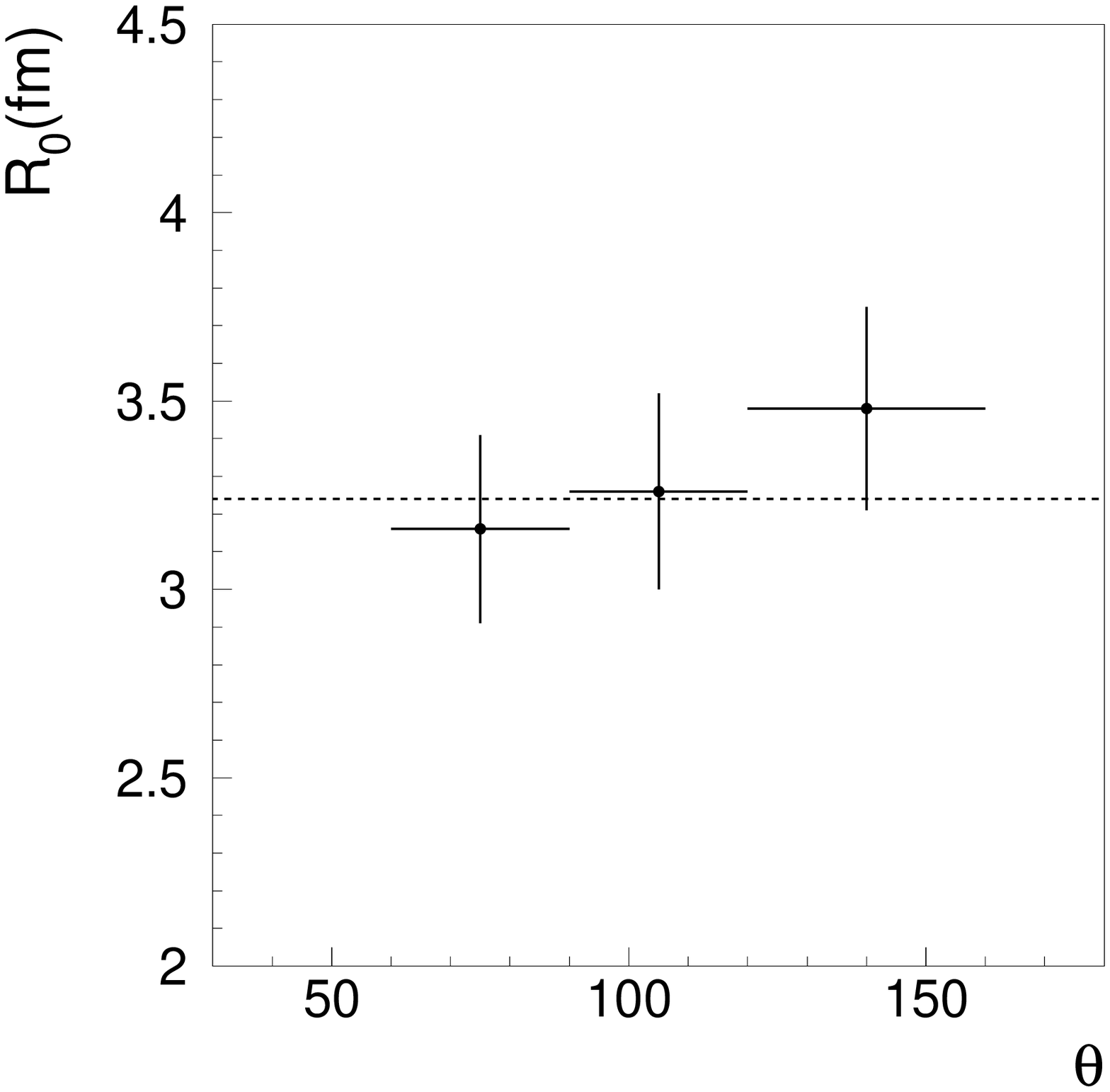} 
\vskip -2.0cm 
\caption[x]{\small Polar angle
dependence of the proton-source radius parameters. The dashed line represents
the weighted average.} \label{p-theta}
\end{figure}

The polar angle (or rapidity) dependence of the proton source radius
parameter was also investigated. Within errors, the two-proton source
radius, $R_0$ showed no rapidity dependence, as seen from Fig.~\ref{p-theta}, over
the entire polar angle interval.

\begin{figure}[t]
\vskip -2.cm
\includegraphics[width=100mm]{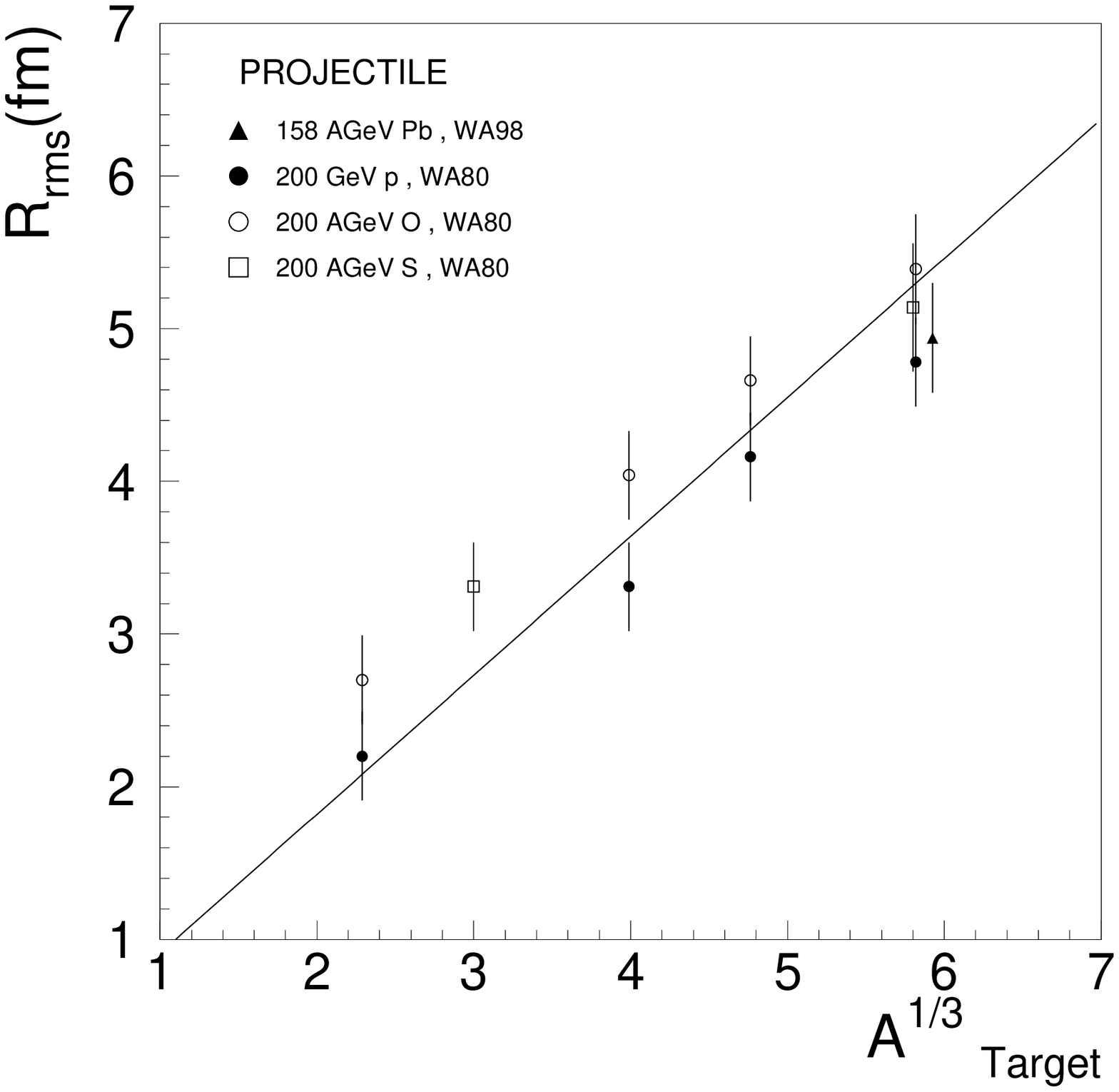} 
\vskip -2.0cm 
\caption[x]{\small The proton
source rms radius parameters as a function of $A_{\rm
Target}^{1/3}$ for proton-nucleus and nucleus-nucleus
collisions at the SPS from this measurement and from
the WA80 experiment~\cite{WA80}. 
The protons have been measured in the target fragmentation
region with -0.6 $< y < $0.6 and 100 MeV/c $ < p_T <$ 650 MeV/c .
The symbols are slightly displaced for the Au and Pb target points 
for clarity. The line represents the fitted function $R_{rms} =
a A_{\rm Target}^{1/3} (a = 0.91 \pm 0.02,\chi^2/n = 1.68 )$. The
WA98 experimental result was obtained using the CRAB code.}
\label{compare-crab}
\end{figure}

Two-proton correlations have also been studied in the target fragmentation 
region at the SPS in the WA80 experiment using the Plastic Ball spectrometer. 
Measurements
were made with 200 $A$ GeV proton, Oxygen, and Sulphur projectiles, 
on targets of C, Al, Ag, and Au~\cite{WA80}. The source radius extracted from the present
analysis of the two-proton correlation for central Pb+Pb 
collisions at 158 $A$ GeV is compared to the radius parameters extracted
in the WA80 experiment in Fig.~\ref{compare-crab}. The radius
parameters are plotted as a function of $A_{\rm Target}^{1/3}$.
All radii have been converted to root-mean-square radii.
Within errors, the radius
parameters measured in the WA80 and WA98 experiments at the SPS
closely follow the target nuclear radius
dependence of $A_{Target}^{1/3}$ with little apparent dependence
on the projectile mass. Since the size of the participant overlap region 
depends on both target and projectile mass, the observation of little
projectile mass dependence indicates that the protons 
detected in the target fragmentation region are emitted from the entire
target volume, following substantial rescattering~\cite{WA80}.

\subsection{Two-deuteron system}

\begin{figure}[ht]
\vskip -2.0cm
 \includegraphics[width=100mm]{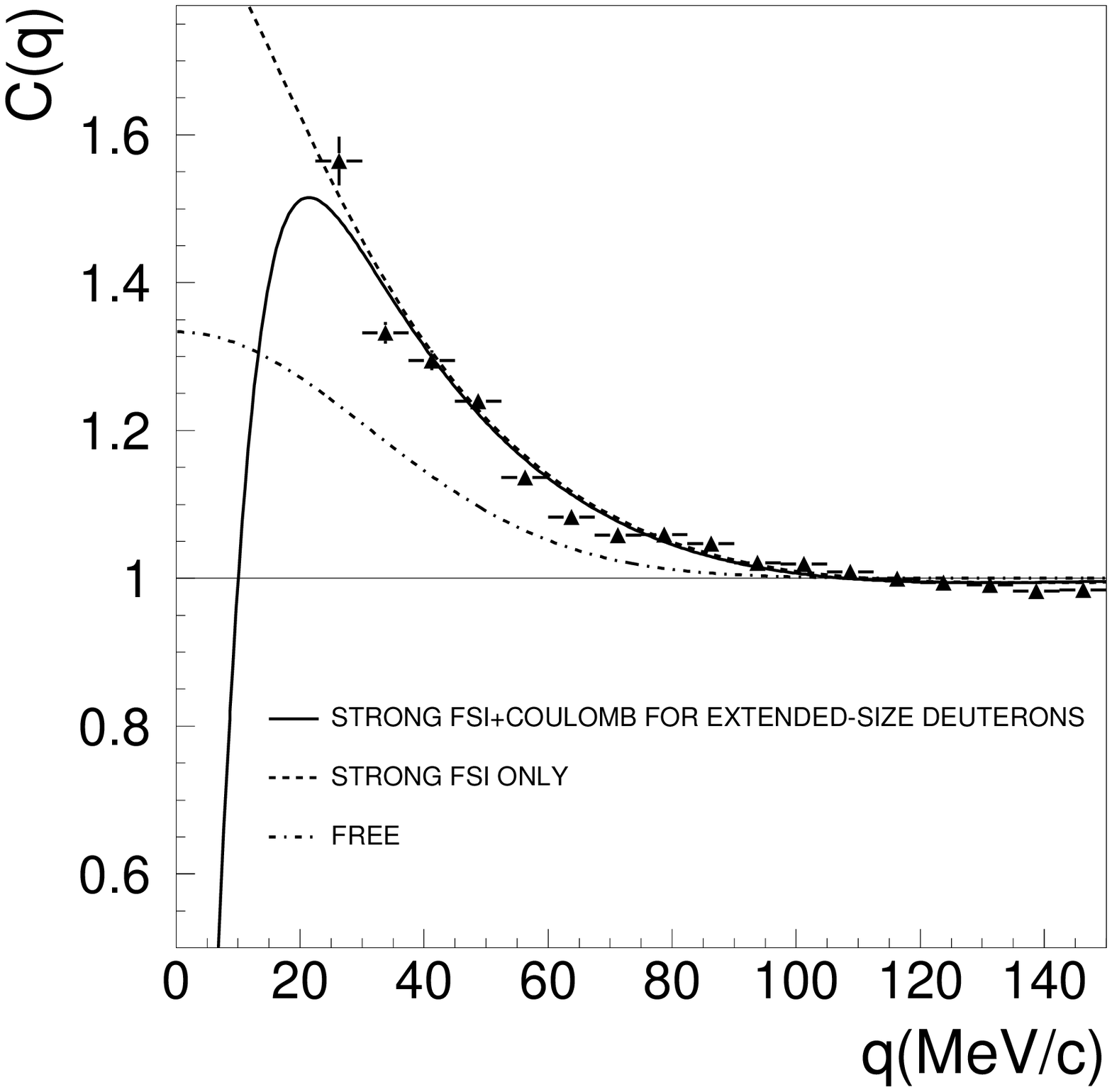} 
\vskip -2.0cm 
 \caption[x]{\small The measured two-deuteron correlation function. The errors shown are statistical only. The solid line
represents the result of the calculation which provides the best
fit to the data with the final state interaction included. The dashed and dot-dashed lines
show the correlation function with only strong final state interaction taken into account 
and that for free propagation, respectively. }
\label{d-corr}
\end{figure}


Figure \ref{d-corr} shows the measured two-deuteron correlation
function and the theoretical calculation which gives the best fit
to the data. The solid and dashed line represent the correlation
functions calculated with final state interaction taken into
account and that for free propagation. The fitted radius parameter
is 1.59$\pm$ 0.02(stat.)$\pm$ 0.20(syst.) fm. The range $R$ and
the strength $s$ for the delta-shell potential are found to be
$0.52\pm 0.01$ fm and $1.08\pm 0.01$, respectively. 
The measured two-deuteron correlation and extracted radius parameters
show no dependence on the polar angle, similar to observations
for the two-proton correlation (see Fig.~\ref{p-theta}).

%

\section{Multi-dimensional analysis}

\begin{table*}[ht]
\caption  {Fitted parameters of the Bertsch-Pratt parametrization of
the two-proton correlation function using the delta-shell model
or CRAB model final state interaction calculations. Errors are statistical + systematic.}
\begin{center}
\begin{tabular}{|c|c|c|c|c|}\hline
& Delta-shell Model & Delta-shell Model & CRAB Model & CRAB Model \\
& (with cross term) &
(without cross term) &  (with cross term) & (without cross term) \\
\hline $\lambda$ &0.76$\pm$0.02$\pm$0.04 & 0.76$\pm$0.02$\pm$0.04
& 0.78$\pm$0.02$\pm$0.04 & 0.79$\pm$0.03$\pm$0.04 \\ \hline
$R_l$(fm) &4.61$\pm$0.06$\pm$0.49 &
4.58$\pm$0.06$\pm$0.47&4.94$\pm$0.07$\pm$0.42 &
4.94$\pm$0.08$\pm$0.39
\\ \hline
$R_s$(fm) &4.43$\pm$0.07$\pm$0.45 &
4.47$\pm$0.08$\pm$0.44&4.81$\pm$0.08$\pm$0.41 &
4.78$\pm$0.08$\pm$0.44 \\ \hline $R_o$(fm) &5.46$\pm$0.10$\pm$0.59
&5.48$\pm$0.09$\pm$0.57&5.64$\pm$0.12$\pm$0.36 &
5.70$\pm$0.14$\pm$0.42 \\ \hline $R_{ol}^2$(fm$^2$)&
-1.65$\pm$0.54$\pm$0.52& &-1.47$\pm$0.49$\pm$0.64 &
\\ \hline
$\chi^2 {\rm /ndf}$ & 1.36 & 1.36 & 1.38 & 1.38 \\
\hline
\end{tabular}
\end{center}
\label{p-bp}
\end{table*}

For consistency of notation with previous analyses, the
momentum difference $\mathbf {Q}=(\mathbf{p}_1-\mathbf {p}_2)$ is
used in the multi-dimensional analysis. In this case the
two-particle correlation $C(Q)$ is given by:
\begin{equation}
 C(Q)=1+\tilde{\rho}(Q)^2.
\label{comp1}
\end{equation}
Note that since the correlation function $C(q)=1+\tilde{\rho}(2q)^2$ was used
for the calculations presented in section 3,
a factor $\sqrt{2}$ should be applied to the radius parameters
obtained from the one-dimensional analysis results presented above
when comparing with the radius 
parameters obtained from the following multi-dimensional analysis.

The multi-dimensional analysis has been performed in the
Longitudinally CoMoving System (LCMS), in which the longitudinal
pair momentum vanishes ($\mathbf p_{z1}+\mathbf p_{z2}=0$).

In order to extract that part of the correlation function of identical
particles that is due to quantum statistical effects only, 
the measured multi-dimensional correlation functions have 
been "corrected" for the effect of the
strong and Coulomb final state interactions (FSI).
The corrected correlation function is defined as:
    \begin{equation}
 C(p_1,p_2) = K\, C_{raw}(p_1,p_2)
\label{fsicorr}
    \end{equation}
where $K=C_{FREE}(Q)/C_{FSI}(Q)$ is the correction factor, and
$C_{FREE}(Q)$ and $C_{FSI}(Q)$ are the calculated correlation functions 
for free propagation and with final state interactions included,
respectively.

\subsection{Two-proton system}

\vskip 0.5cm \noindent {\bf The Bertsch-Pratt (BP)
parametrization} \vskip 0.5cm

 In the Bertsch-Pratt
parameterization~\cite{PRATT84,BERTSCH}, the correlation function for two fermions has the form:
\begin{eqnarray}
\label{BP}
 C = N \left( 1- \lambda \E^{- Q_l^2 \cdot R_l^2 - Q_s^2 \cdot R_s^2
 - Q_o^2 \cdot R_o^2 - 2 \cdot Q_o Q_l \cdot R_{ol}^2 } \right)
\end{eqnarray}
where $N$ is the normalization constant and $0 \leq \lambda \leq 1$ is the
strength of the correlation which depends on the chaoticity of the emitter and on
the purity of the measured data sample. The momentum difference vector $\mathbf{Q}$
of two particles is decomposed into a longitudinal component along
the beam axis $(Q_l)$ and a component transverse to the beam axis
$(Q_T)$. The transverse momentum difference is further decomposed
into the outward and sideward components. The outward and sideward
momentum differences $(Q_o)$ and $(Q_s)$ are defined as the
components of the transverse momentum difference parallel and
perpendicular to the total transverse momentum of the pair,
respectively.
The sign convention is that $Q_s$ is always
positive, $Q_l$ and $Q_o$ are allowed to be positive or negative.
$R_{ol}^2$ is the "out-longitudinal" cross term \cite{CROSSTERM}. 
Its value can be either positive or negative.
In the LCMS system the average resolutions in the $Q_s$, $Q_l$, and $Q_o$ 
components for the pp system are 20, 23, and 24 MeV/c, respectively.

\begin{figure}[ht]
\vskip -2.0cm
\includegraphics[width=100mm]{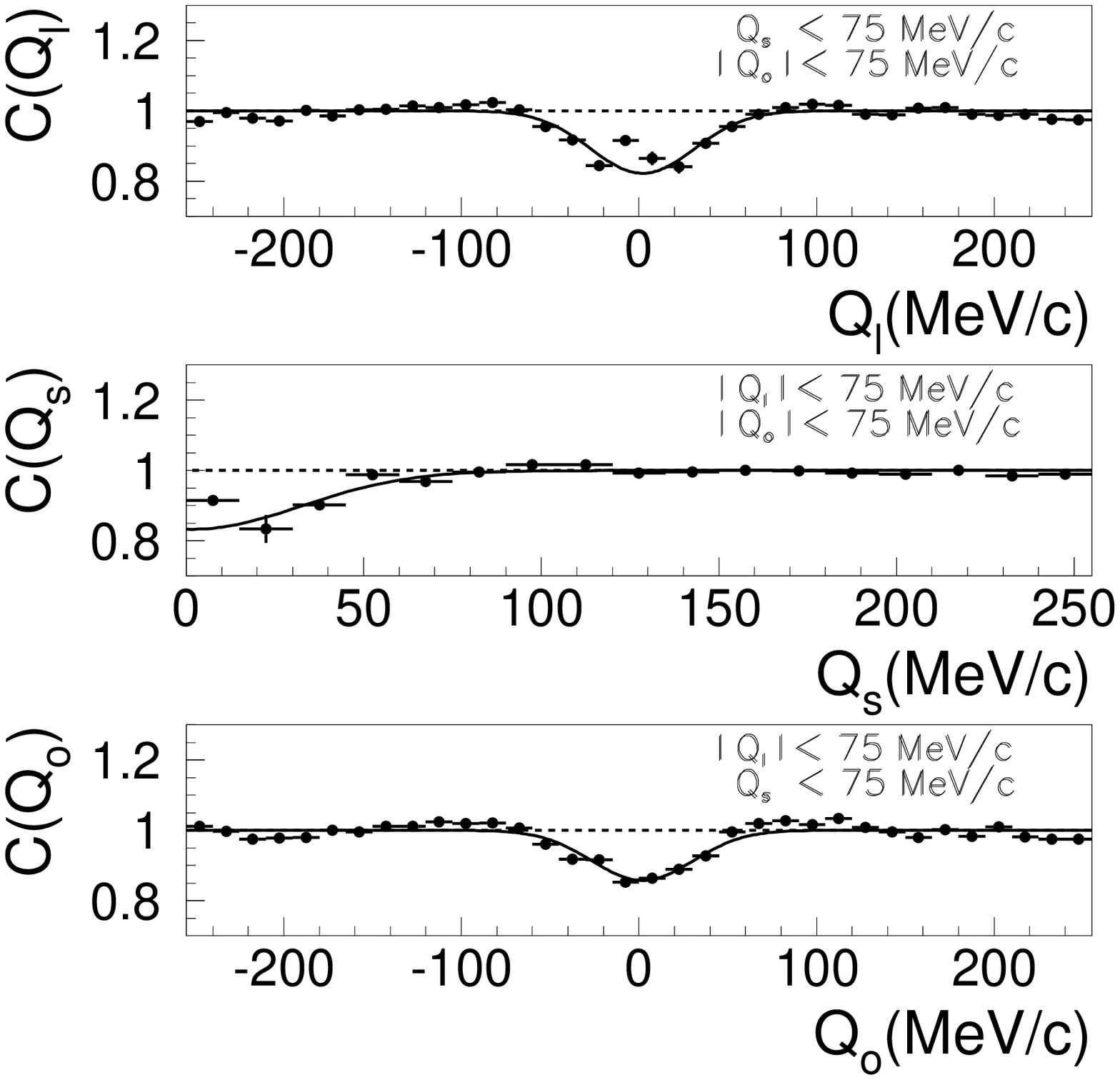} 
\vskip -2.0cm
 \caption[x]{\small Projections of the measured two-proton
correlation function corrected with the CRAB calculation 
for the FSI. The solid lines show the projections of the
fit with the Bertsch-Pratt parametrization. Projections for each $Q$
component, for both the experimental data and the fit, have been
integrated up to 75 MeV/c over the other two $|Q|$ components. The
errors shown are statistical only. } \label{p-projection}
\end{figure}

Figure \ref{p-projection} shows projections of the measured two-proton correlation
functions, corrected for final state interactions (Eq.~(\ref{fsicorr})),  
onto the $Q_l$, $Q_s$, and $Q_o$ axes. Also shown  are the corresponding
projections of the fit of Eq.~(\ref{BP}), shown as solid
lines. The projected results and the
fit results for a given $Q$ component have been integrated up to 75 MeV/c over
the other two $|Q|$ components.

The parameters extracted from  fits to the data using either the
delta-shell corrected and CRAB corrected correlation functions are
presented in Table~\ref{p-bp}.  Also listed in this table are the
fitted parameters without the cross term $R_{ol}^2$. Within 
errors, the delta-shell and CRAB corrections lead to the
same values of fitted parameters.  The fitted radius parameters
with and without the cross term do not differ within errors.
The cross term $R_{ol}$ vanishes in the LCMS frame
for longitudinally {\it{boost-invariant}} systems.
The observation that the cross-term $R_{ol}^2$ is small, indicates that the proton
source undergoes an approximate boost invariant expansion.

\begin{figure}[ht]
\vskip -2.0cm
\includegraphics[width=100mm]{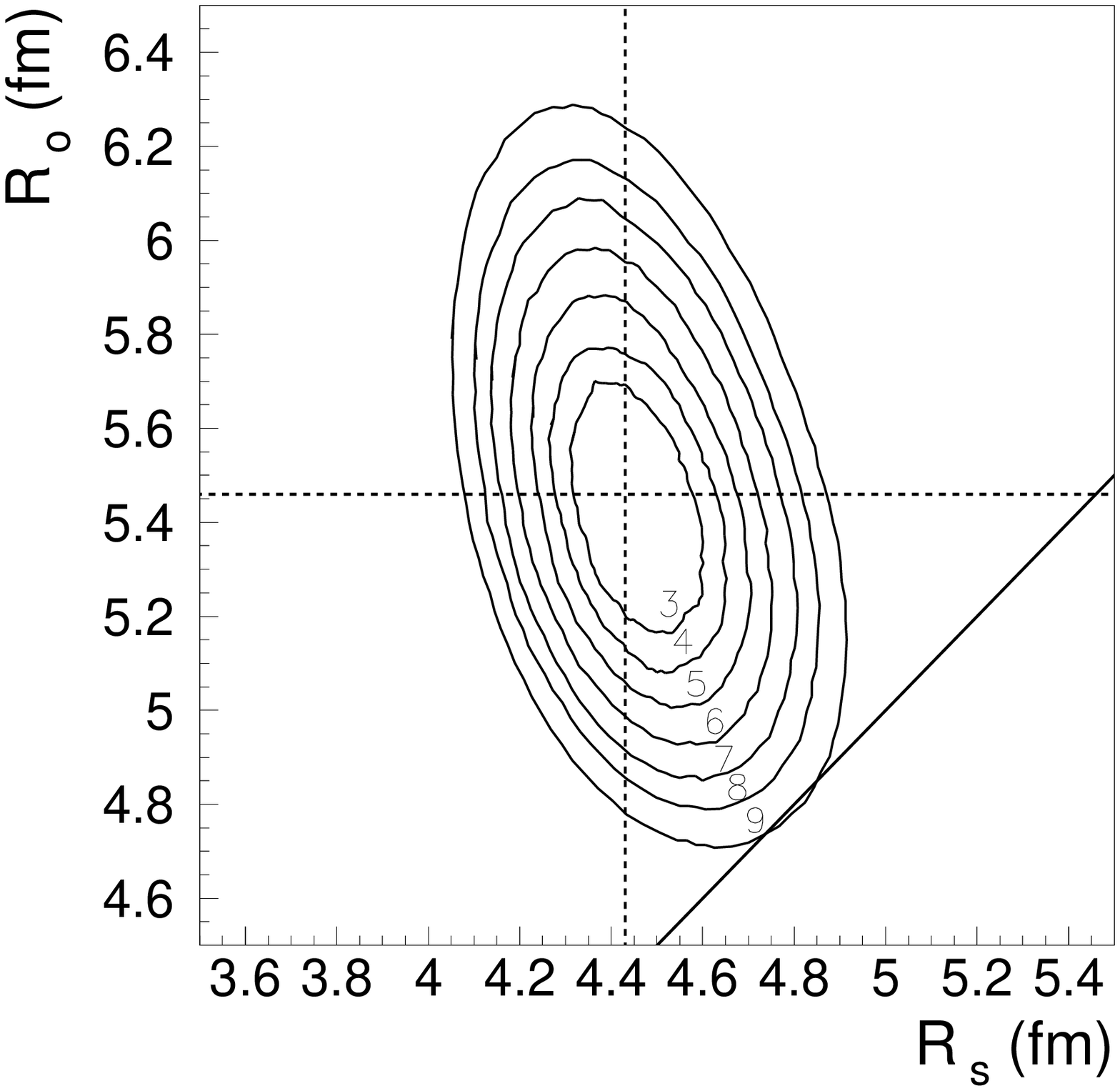} 
\vskip -2.0cm
 \caption[x]{\small Contour lines for $R_o$ versus $R_s$. 
The labels $3,4,...,9$ refer to the number of standard deviations
from the minimum $\chi^2$. The solid line corresponds to $R_o=R_s$. }
\label{contour}
\end{figure}

The statistical significance of the fit is demonstrated in 
Fig.~\ref{contour}, where contour lines of equal confidence level
for $R_o$ versus $R_s$ are plotted. The spacing of contours
corresponds to one standard deviation in $\chi^2$ value. The
solid line, which represents equal sideward and outward radii, is
9 standard deviations from the minimum $\chi^2$ value.

For transparent sources, where particles are emitted from throughout the entire
volume,  the longitudinal radius $R_l$ is interpreted to measure the
longitudinal size of the system directly, while the sideward radius
provides a measurement of the transverse size. The duration of emission
$\Delta \tau$ is related~\cite{P,B} to the sideward radius
$R_s$ and outward radius $R_o$ as:
\begin{eqnarray}
\label{time2} 
\Delta\tau = {1\over \beta_T}\sqrt{R_o^2-R_s^2},
\end{eqnarray}
where $\beta_T$ is the transverse velocity of the particle pair. Notice that 
Eq.~(\ref{time2}) is valid only when opacity
effects and transverse flow are ignored. The excess of $R_o$ over
$R_s$ is due to the duration of emission from the source in which
particles with transverse velocity $\beta_T$ travel a distance
$\beta_T\Delta\tau$ towards the detector, resulting in an apparent
extension of the source in that direction. The sideward radius
parameter $R_s$ is not affected by the duration of emission since it is
perpendicular to $\beta_T$.  

Within errors, the radius parameters ($R_l,R_s,R_o$)
of the BP fit are quite close to each other (Table~\ref{p-bp})
which suggests a volume emission of the protons. The extracted 
$R_o$ and $R_s$ parameters, and the measured average value 
of $\beta_T=0.278\pm 0.09$ for the proton pairs, 
gives $11.5\pm 2.5$ fm/c as an estimate of the emission duration.

In the case of a Bjorken scenario, with longitudinal expansion
of the system without transverse flow, the
$m_T$ dependence of the longitudinal radius reflects the lifetime of
the source\cite{MS}:
\begin{eqnarray}
\label{lifetime} 
R_l=\tau_f(T_f/m_T)^{1/2},
\end{eqnarray}
where $T_f$ is the freeze-out temperature and
$m_T=\sqrt{m^2+p_T^2}$. Under the assumption that $T_f=120$ MeV,
a source lifetime of about 13 fm/c is obtained.

Strictly speaking, the interpretation of $\tau_f$ as the
freeze-out time is based on the assumption that the longitudinal
boost-invariant velocity profile existed not only at the time of
freeze-out but also throughout the dynamical evolution of the
reaction zone. If this were not the case, then $\tau_f$ would be larger.
Hence, $\tau_f$ estimated by Eq.~(\ref{lifetime}) gives a {\em lower} limit
for the lifetime of the source \cite{wiedeman}.

\vskip 0.5cm \noindent {\bf The Yano-Koonin-Podgoretsky (YKP)
parametrization} \vskip 0.5cm

 In the Yano-Koonin-Podgoretsky parameterization\cite{YKP,Misha},
the two-proton correlation function is defined by the formula:
\begin{eqnarray}
\label{YKP1}
 C = N \left( 1- \lambda \E^{- Q_T^2 R_T^2 -
(Q_l^2-Q_0^2)R_l^2-(Q\cdot U)^2(R_0^2+R_l^2) }  \right)
\end{eqnarray}
where $U=\gamma(1,0,0,\beta)$ is a 4-velocity with only a
longitudinal component, $\gamma=1/\sqrt{1-\beta^2}$, $Q_T$ and
$Q_l$ are the components of the momentum difference projected 
onto the transverse and
longitudinal directions, respectively, $Q_0$ is the 
difference in energies. The sign convention is $Q_0$ is always
positive and $Q_l$ can be either positive or negative. This
parametrization has an advantage in that
the three YKP radius parameters do not depend on the longitudinal
velocity of the measurement frame, and the velocity $\beta$ is
closely related to the velocity of the effective particle emitter.

The parameters extracted from YKP fits to the delta-shell
corrected or CRAB corrected correlation functions are listed in
Table~\ref{p-ykp}. Within errors, the fitted
parameters obtained with the two different final state interaction
corrections are consistent with each other.

\begin{table}[ht]
\caption  {Fitted parameters of the YKP parametrization of the
two-proton correlation functions using either the delta-shell model or
CRAB model FSI corrections. Errors are statistical + systematic. }
\begin{center}
\begin{tabular}{|c|c|c|}\hline
& Delta-shell Model & CRAB Model \\ \hline $\lambda$
&0.75$\pm$0.04$\pm$0.04 & 0.77$\pm$0.03$\pm$0.03 \\ \hline
$R_l$(fm) & 4.64$\pm$0.16$\pm$0.36& 4.97$\pm$0.08$\pm$0.46 \\
\hline $R_T$(fm) &  4.42$\pm$0.13$\pm$0.40& 4.73$\pm$0.06$\pm$0.46
\\ \hline
 $R_0$(fm) & 10.54$\pm$0.46$\pm$1.18& 10.58$\pm$0.46$\pm$1.12 \\
 \hline
 $\beta$& 0.032$\pm$0.023$\pm$0.010& 0.036$\pm$0.014$\pm$0.011 \\
 \hline
 $\chi^2 {\rm /ndf}$ & 1.42 & 1.47 \\ \hline
\end{tabular}
\end{center}
\label{p-ykp}
\end{table} 

Since the YKP velocity $\beta$ is close to 0, it demonstrates that the LCMS
frame coincides with the YK frame (the frame for which the YKP velocity
parameter $\beta$ vanishes). For {\em transparent sources}, 
and in the absence of flow,
the radius parameters in the YK frame have a convenient physical
interpretation. $R_l$ and $R_T$ are then interpreted as measures of the
longitudinal and transverse size of the source, while $R_0$
reflects directly the duration of emission from the source.

With similar fitted $R_l$ and $R_T$ parameters, and 
longitudinal velocity $\beta$ compatible with 0, the YKP fit
indicates a symmetric emission geometry with approximate boost invariant
expansion of the source. The emission duration extracted from the 
YKP or BP fits are in agreement within errors.

The fitted YKP parameters are seen to be consistent with the BP parameters.
Since the BP and YKP parametrizations are mathematically
equivalent and differ only in the choice of independent components of
q, the agreement between the two sets of fitted parameters serves 
additionally as a consistency check of the fitting procedures.

\subsection{Two-deuteron system}

\vskip 0.5 cm\noindent{\bf The Gaussian parametrization in the LCMS}\vskip 0.5 cm

\begin{figure}[ht]
 \vskip -2.0cm
 \includegraphics[width=90mm]{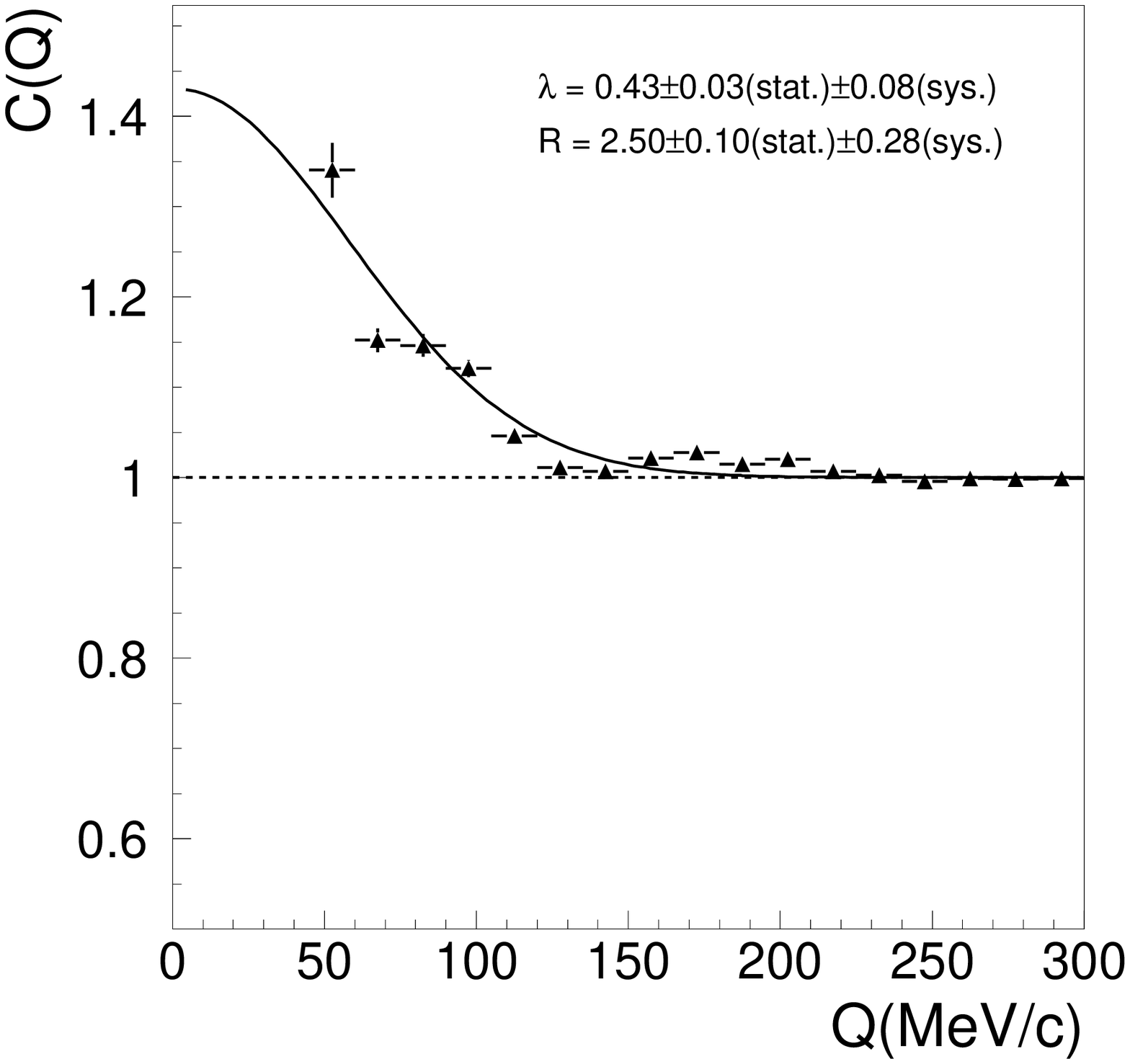} 
\vskip -2.0cm
\caption[x]{\small The fitted two-deuteron correlation function as a
 function of momentum difference $Q$ in the LCMS frame after correction for final
 state interactions.} \label{q-lcms}
\end{figure}

Fig. \ref{q-lcms} shows the experimental one-dimensional two-deuteron correlation function 
in the LCMS frame, after correction for final state interactions using the delta-shell 
model described in section 3. The correlation is fitted to the Gaussian correlation function:

\begin{equation}
C(Q)=N \left(1+\lambda \E^{- Q^2 R^2} \right).
\end{equation}

The fitted parameters are: $\lambda
=0.43\pm0.03$(stat.)$\pm0.08$(syst.),
$R=2.50\pm0.10$(stat.)$\pm0.28$(syst.) fm.
The radius parameter is in agreement within errors with the
radius ($R=1.59\times\sqrt{2}=2.25$ fm) extracted by fit to the
one-dimensional correlation function with the full delta-shell model
calculation (see Fig.~\ref{d-corr}).

\vskip 0.5 cm\noindent{\bf The Bertsch-Pratt parametrization}\vskip 0.5 cm

The two-deuteron correlation function has been corrected by the 
calculated delta-shell model two-deuteron final state interaction 
and fitted with the 
two-boson correlation function using the Bertsch-Pratt 
parametrization:
\begin{eqnarray}
\label{BP1}
 C =N \left( 1+ \lambda \E^{- Q_l^2 \cdot R_l^2 - Q_s^2 \cdot R_s^2
 - Q_o^2 \cdot Q_o^2 - 2 \cdot Q_o Q_l \cdot R_{ol}^2 } \right).
\end{eqnarray}

\begin{figure}[ht]
\vskip -2.0cm
\includegraphics[width=100mm]{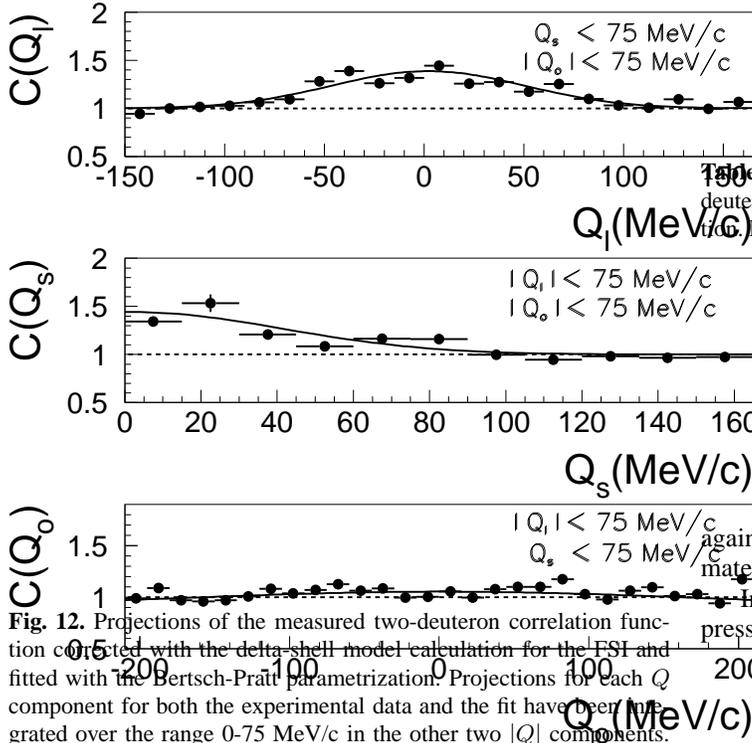} 
\vskip -2.0cm
 \caption[x]{\small Projections of the measured two-deuteron
correlation function corrected with the delta-shell model calculation
for the FSI and fitted
with the Bertsch-Pratt parametrization. Projections
for each $Q$ component for both the experimental data and the fit
have been integrated over the range 0-75 MeV/c in the other two $|Q|$
components. The errors shown are statistical only. }
\label{d-projection}
\end{figure}

Projections of the measured two-deuteron correlation function on the
$Q_l$, $Q_s$, and $Q_o$ axes are shown in 
Fig.~\ref{d-projection}. The measured and fitted projections in a given $Q$
component have been integrated up to 75 MeV/c in the other two $|Q|$
components. All $Q$ components have approximately the same average
resolution of 32 MeV/c.

The fitted parameters, with and without the cross term $R_{ol}^2$,
are listed in Table~\ref{d-bp}. Within errors, the
values of parameters extracted from the two different fits are
the same. Similar to the result for the proton source, the small cross-term  indicates that
the deuteron source also experienced an approximate boost
invariant expansion.

\begin{table}[ht]
\caption  {Fitted parameters of the Bertsch-Pratt parametrization of
the two-deuteron correlation functions using the delta-shell model
FSI correction. Errors are statistical + systematic. }
\begin{center}
\begin{tabular}{|c|c|c|}\hline
& with cross term  & without cross term \\ \hline
 $\lambda$ &0.37$\pm$0.02$\pm$0.19&  0.37$\pm$0.02$\pm$0.19 \\
 \hline
$R_l$(fm) & 2.17$\pm$0.10$\pm$0.24& 2.13$\pm$0.08$\pm$0.18 \\
\hline  $R_s$(fm)& 3.14$\pm$0.11$\pm$0.78& 3.10$\pm$0.10$\pm$0.67
\\ \hline
 $R_o$(fm) & 0.01$\pm$0.08$\pm$0.02 &0.01$\pm$0.04$\pm$0.03 \\
\hline  $R_{ol}^2$(fm$^2$)& 0.20$\pm $ 0.08 $\pm$0.69& \\ \hline
$\chi^2 {\rm /ndf}$ & 1.52 & 1.52 \\ \hline
\end{tabular}
\end{center}
\label{d-bp}
\end{table}

%

Within errors, the fitted  $R_l$ and $R_s$
parameters are close to the radius value obtained from the one-dimensional
fit, while $R_o$ is consistent with 0. For a transparent, azimuthally symmetric
emission source, it is expected that $R_s$ and $R_o$ should be similar, except that
$R_o$ will be extended by the duration of emission, as discussed with respect to
Eq.~(\ref{time2}). The observation that $R_o<<R_s$ is inconsistent with 
such a transparent source, but may result naturally from surface emission from an 
opaque source~\cite{opaque1,opaque2}. 
Since deuterons are relatively large weakly bound objects, 
it might be expected that only those deuterons produced on the
freeze-out surface via coalescence of an emitted proton and neutron
survive. This would naturally lead to an opaque emission source,
as observed.


\vskip 0.5cm \noindent {\bf The Yano-Koonin-Podgoretsky
parametrization} \vskip 0.5cm

 The two-boson correlation function in the Yano-Koonin-Podgoretsky parametrization is given by:
\begin{eqnarray}
\label{YKP2}
 C = N \left(1+ \lambda \E^{- Q_T^2 R_T^2 -
(Q_l^2-Q_0^2)R_l^2-(Q\cdot U)^2(R_0^2+R_l^2)} \right).
\end{eqnarray}

\begin{table}[ht]
\caption  {Fitted parameters of the YKP parametrization of the
two-deuteron correlation functions using the delta-shell model
FSI correction. Errors are statistical + systematic. }
\begin{center}
\begin{tabular}{|c|c|}\hline
$\lambda$ &0.29$\pm$0.02$\pm$0.09 \\ \hline
$R_l$(fm) &  1.29$\pm$0.07$\pm$0.42 \\
\hline $R_T$(fm) & 2.35$\pm$0.05$\pm$0.39 \\ \hline
 $R_0^2$(fm$^2$) & -46.95$\pm$1.69$\pm$8.82 \\ \hline
 $\beta$& 0.020$\pm$0.002$\pm$0.007 \\ \hline
 $\chi^2 {\rm /ndf}$ & 1.42 \\ \hline
\end{tabular}
\end{center}
\label{dd-ykp}
\end{table}

The parameters of the YKP fit to the two-deuteron correlation
are listed in Table~\ref{dd-ykp}. The
longitudinal velocity $\beta$ is close to 0 indicating that the
LCMS coincides with the YK frame and again indicates that 
the deuteron source undergoes an
approximate boost invariant expansion.

In the YK frame the YKP radius parameter $R_0$ can be expressed as \cite{WHTW96}:
\begin{equation}
R_0^2= \frac{1}{\beta_\perp^2} \left( R_o^2 - R_s^2 \right),
 \label{life-ykp}
\end{equation}
where $\beta_\perp$ is the velocity of the particle pair transverse 
to the beam direction, and $R_o$ and $R_s$ are the outward and
sideward radii of the BP parametrization. 
The negative value of $R_0^2\approx
 -47$ fm$^2$ obtained from the YKP fit is in agreement with the observation
of strong opacity~\cite{wiedeman,opaque1,tomasik} of the deuteron source  
made in the previous section.
%

\section{Discussion}

Although one does not necessarily expect boost-invariance in 
the target fragmetnation region at SPS energies, the small
values of the cross-term $R_{ol}$ in the BP parametrization as well 
as the consistency with zero longitudinal velocity $\beta$ in the YKP 
parametrization of the correlation functions, indicates that both proton and 
deuteron sources exhibit boost-invariant expansion.

The measured one-dimensional radius parameter extracted for 
protons is markedly larger than
that for deuterons. Comparison of the present results for protons and deuterons
with those for $\pi^+$, also measured in the Plastic Ball~\cite{karpio},
and $\pi^-$ measured in the WA98 negative-charged particle 
spectrometer~\cite{wa98-pion,wa98-2pion-new},
and with $K^+$ measured in the NA44 experiment~\cite{kohama}
for central Pb+Pb collisions at 158 $A$ GeV, reveals that the radii are observed so show a mass
ordering with 
 $R_{\pi\pi}> R_{\rm KK}>R_{\rm pp}>R_{\rm dd}$. 
This observation is in agreement with previous reports \cite{ALEXANDER} that
lighter particles tend to give larger radius parameters.
It is interesting that a large loosely bound composite object 
like the deuterons follows the same trend as observed 
for elementary particles. 
It has been shown by
Alexander \cite{ALEXANDER,ALEXANDER2} that 
the source-size mass dependence of hadrons emerging
from ${Z^0}$ decays produced in $e^+e^-$ annihilation
can be reproduced using the Heisenberg
uncertainty relations to derive the relations:
\begin{equation}
\label{mass-fit}
 R =  \frac{A}{\sqrt{m}},
\end{equation}
and
\begin{equation}
\label{mass-fit1}
 R_l = \frac{A}{\sqrt{m_T}},
\end{equation}
where $A = c\sqrt{\hbar \Delta t}$ is a time scale constant, $R_l$
is the longitudinal radius parameter, and $m$ and $m_T$ are the 
mass and the mean transverse mass, respectively. 
An alternative explanation using a QCD derived
potential~\cite{ALEXANDER} proved to be equally successful. 
A different approach to explain the $R$ and $R_l$ mass dependence
was given by Bialas and Zalewski~\cite{BIALAS,BIALAS1,BIALAS2}. In
this description the radius parameter of the source is mass independent
and its apparent decrease is a consequence of the
momentum-position correlation expressed in the
Bjorken-Gottfried condition~\cite{GOTTFRIED,BJORKEN}. However, 
a study of purely kinematical considerations~\cite{SMITH} led to the
conclusion that this is unlikely to account for the observed
$R(m)$ dependence.

\begin{figure}[ht]
\vskip -2.0cm
\includegraphics[width=90mm]{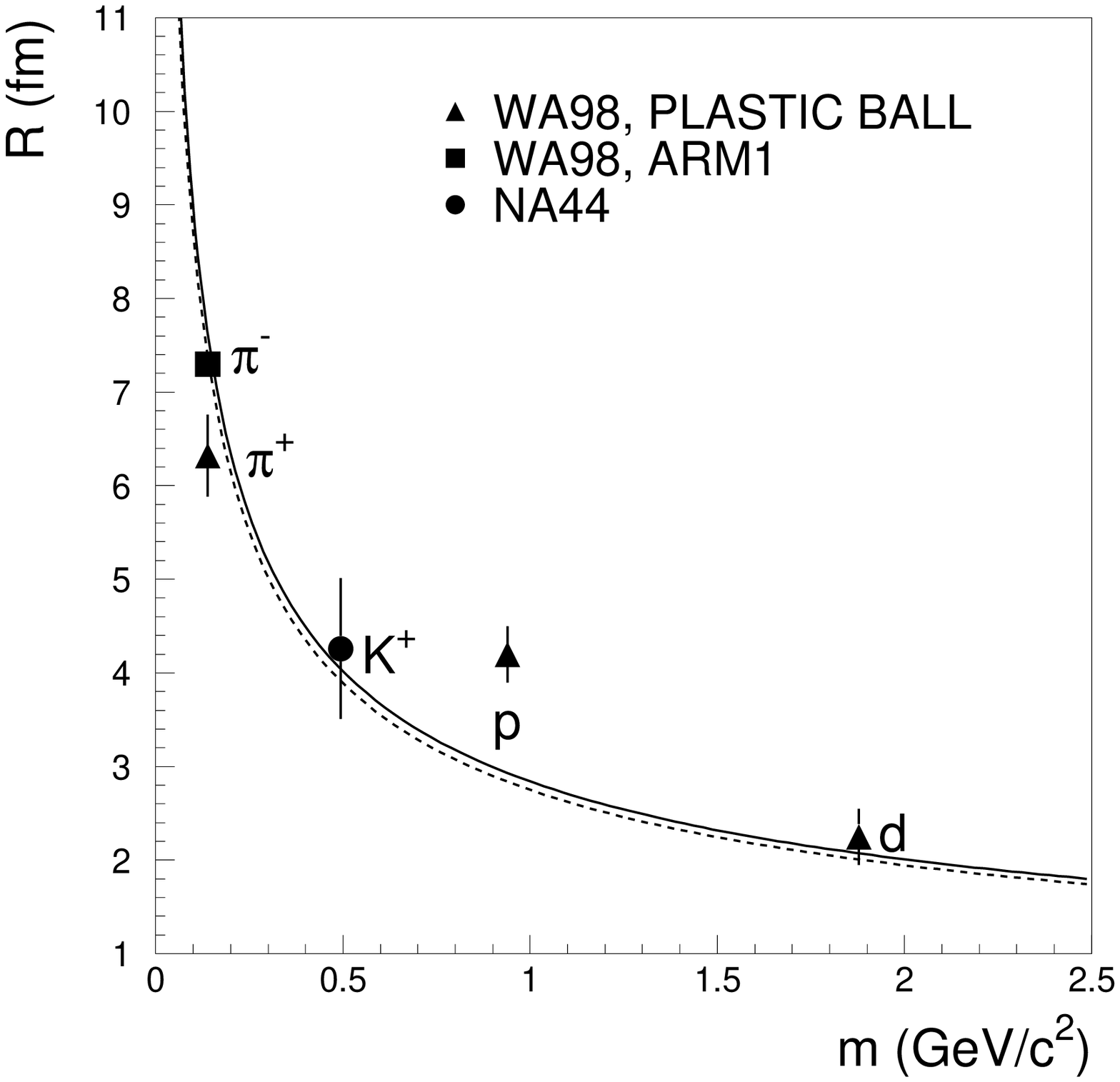} 
\vskip -2.0cm
 \caption[x]{\small The dependence of radius parameters extracted from
 one-dimensional two-particle correlations on the particle mass in 
 central Pb+Pb collisions at 158 $A$ GeV.
 The error bars represent the statistical and systematic errors added in quadrature.
 The solid curve is the result of fit of Eq.~(\ref{mass-fit}) to the three
WA98 data points
 measured with Plastic Ball spectrometer. The dashed curve
 includes also the $\pi^-$ data from WA98 and kaon data
 from the NA44 experiment.} \label{mass}
\end{figure}

Radius parameters  extracted from one-dimensional 
two-particle correlation functions 
for central Pb+Pb collisions at 158 $A$ GeV are shown
as a function of the particle mass in Fig.~\ref{mass}. 
The $\pi^-\pi^-$ radius parameter was measured with the
negative-charged particle
spectrometer of the WA98 experiment~\cite{wa98-pion,wa98-2pion-new} and the two-kaon
radius parameter was reported by
the NA44 Collaboration~\cite{kohama},  both measured near
mid-rapidity $\langle y\rangle\sim 2.9$. In general, the radii
follow the $\sqrt{m}$ dependence expectation of Eq.~(\ref{mass-fit}).
A fit of Eq.~(\ref{mass-fit}) 
to the $\pi^+$~\cite{karpio}, proton, and deuteron radii measured with the 
WA98 Plastic Ball gives $A= 2.84\pm 0.13$ fm GeV$^{1/2}$ with
$\chi^2$/ndf = 12, as shown by the solid curve. This corresponds
to a time scale of $\Delta t =40.8$ fm/c using the above expression for
$A$, or to a freeze-out lifetime of $\tau_f =8.2$ fm/c using Eq.~(\ref{lifetime}). A fit to all
data points (dashed line) results in $A= 2.75\pm 0.04$ fm
GeV$^{1/2}$ with $\chi^2$/ndf = 7.2.

\begin{figure}[ht]
\vskip -2.0cm
\includegraphics[width=90mm]{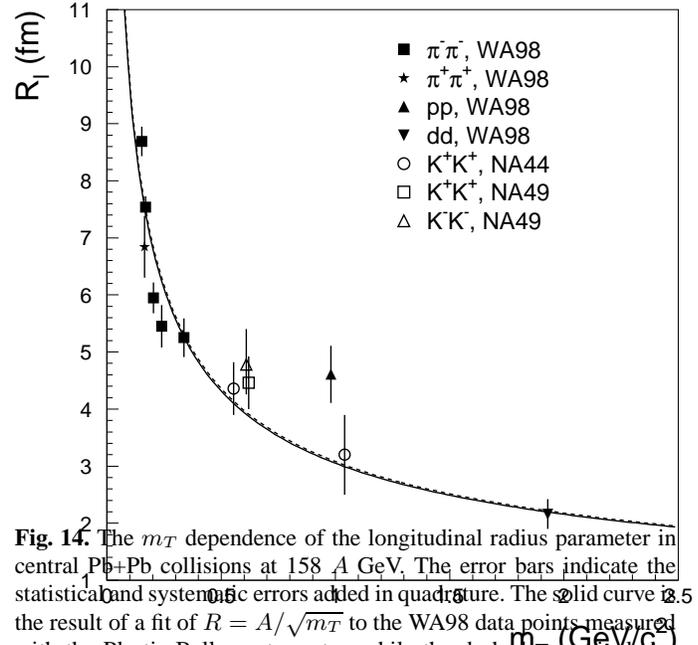} 
\vskip -2.0cm
\caption[x]{\small The $m_T$ dependence of the longitudinal radius 
parameter in central Pb+Pb collisions at 158 $A$ GeV.
 The error bars indicate the statistical and systematic errors added in quadrature.
 The solid curve is the result of a fit of $R=A/\sqrt{m_T}$ to the WA98 data points 
 measured with the Plastic Ball
 spectrometer, while the dashed
 curve includes also mid-rapidity $\pi^-$ data from WA98 and kaon data from the 
 NA44 and NA49 experiments. }
\label{mt-rl}
\end{figure}

\begin{figure}[ht]
\vskip -2.0cm
\includegraphics[width=90mm]{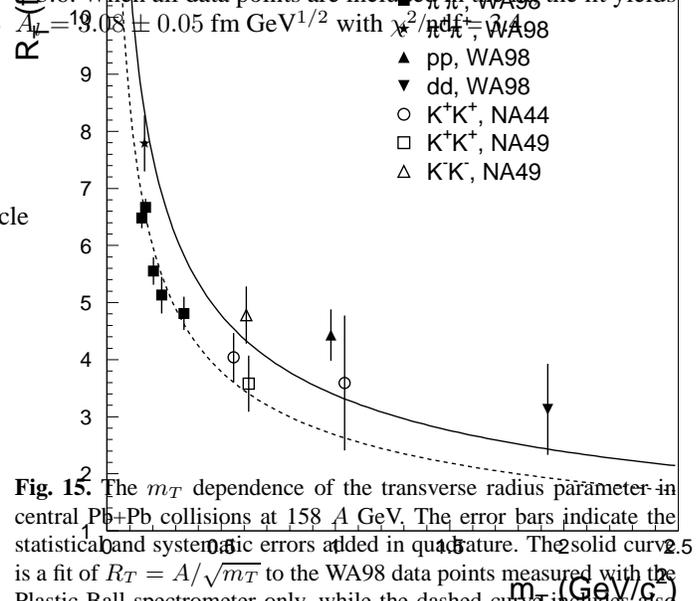} 
\vskip -2.0cm
 \caption[x]{\small The $m_T$ dependence of the transverse radius parameter in 
 central Pb+Pb collisions at 158 $A$ GeV.
 The error bars indicate the statistical and systematic errors added in quadrature.
The solid curve is a fit of $R_T=A/\sqrt{m_T}$ to the WA98
data points measured with the Plastic Ball
 spectrometer only, while the dashed
 curve  includes  also mid-rapidity $\pi^-$ data from WA98 and kaon data from the 
 NA44 and NA49 experiments. }
\label{mt-rs}
\end{figure}

Previous results on pion and kaon interferometry for S+Pb and
Pb+Pb reactions at the SPS have demonstrated scaling of the $R_l$ 
and $R_T$ radius parameters on the transverse mass, 
$m_T$~\cite{na44-pion,na44-kaon}.
The longitudinal and transverse radius parameters from the multi-dimensional
YKP analysis of the two-proton and two-deuteron correlation functions
are plotted as a function of the mean transverse mass in 
Figs.~\ref{mt-rl} and \ref{mt-rs}. Also shown are
radius parameters for $\pi^+$ measured with the Plastic 
Ball~\cite{karpio}. Radii measured near mid-rapidity for $\pi^-$ in 
WA98~\cite{wa98-pion,wa98-2pion-new}, and for kaons from the NA44~\cite{na44-kaon} 
and NA49~\cite{na49-kaon} experiments are also shown. 
The data are compared to fits of the form $R_i = A_i/\sqrt{m_T}$, where $i=l,T$.
As  can be seen from Figs.~\ref{mt-rl} and
\ref{mt-rs}, within experimental uncertainties, the fitted
$R_l$ and $R_T$ radius parameters are consistent with
$A/\sqrt{m_T}$ scaling. A fit to the three WA98 data points
measured near mid-rapidity with the Plastic Ball spectrometer 
yields $A_l= 3.05\pm 0.17$
fm GeV$^{1/2}$ with $\chi^2$/ndf = 3.6. When all data points are
included in the fit, the fit yields $A_l= 3.08\pm
0.05$ fm GeV$^{1/2}$ with $\chi^2$/ndf = 3.4.

Although there is no theoretical 
justification to expect an $R_T$=$A/\sqrt{m_T}$ dependence, the data 
are seen (Fig.~\ref{mt-rs}) to follow such a dependence, within errors.
A fit to the three WA98 data points
measured near mid-rapidity with the Plastic Ball spectrometer 
yields $A_T= 3.38\pm 0.18$~fm~
GeV$^{1/2}$ with $\chi^2$/ndf = 2.4
When all data points are
included in the fit, the fit yields $A_T= 2.68\pm
0.04$ fm GeV$^{1/2}$ with $\chi^2$/ndf = 3.3.
A fit to the pion and kaon data from NA44
experiment alone gave $A_T= 3.0\pm
0.2$~fm~GeV$^{1/2}$ \cite{na44-kaon}.  

\section{Summary}
\label{secsumm}

 In summary, we have measured two-proton and two-deuteron
correlation functions in the target fragmentation region for central
Pb+Pb collisions at 158 $A$ GeV.

In the one-dimensional analysis, the radius parameters
were extracted from the measured correlations using
calculations that assume a static Gaussian-shaped source. The extracted
proton source radius parameter is about two times larger than that of 
the deuteron source. Comparison of the proton and deuteron radius 
parameters with those extracted from one-dimensional two-pion and two-kaon correlations 
reveals a common $A/\sqrt{m}$ dependence with  $A\approx 3$ fm~GeV$^{1/2}$. Such a
dependence can be explained as a consequence of the Heisenberg 
uncertainty relations, under the assumption of a common duration of emission.

The multi-dimensional analysis demonstrates that the 
proton source exhibits a volume emission with long emission time, whereas the
deuteron source is strongly opaque. The cross-term
$R_{ol}^2$ from the Bertsch-Pratt parametrization fit and the longitudinal velocity $\beta$
from the Yano-Koonin-Podgoretsky parametrization fit are consistent with zero, suggesting that the
proton and deuteron sources undergo an approximate longitudinal boost invariant
expansion.

The longitudinal and transverse radius parameters, $R_l$ and $R_T$,
extracted from the multi-dimensional correlation analysis 
follow a common $A/\sqrt{m_T}$ scaling for pions, kaons, protons, and deuterons,
with $A\approx 3$ fm~GeV$^{1/2}$ in both cases. 
The existence of a universal function describing the 
dependence of the radius parameters on the transverse mass for 
different particle species in the mass interval from pion to deuteron 
may indicate an approximately simultaneous freeze-out of the studied hadrons.


We wish to express our gratitude to Professor A. Deloff for supplying his codes for the pp and dd FSI
calculations, and for his guidance and helpful discussions.
We wish to thank the CERN accelerator division for the excellent performance of the SPS
accelerator complex. We acknowledge with appreciation the effort of all engineers,
technicians, and support staff who have participated in the construction of this experiment.
This work was supported jointly by
the German BMBF and DFG,
the U.S. DOE,
the Swedish NFR and FRN,
the Dutch Stichting FOM,
the Polish MEiN under Contract No. 1P03B02230 and CERN/88/2006
The Grant Agency of the Czech Republic under contract No. 202/95/0217,
the Department of Atomic Energy,
the Department of Science and Technology,
the Council of Scientific and Industrial Research and
the University Grants
Commission of the Government of India,
the Indo-FRG Exchange Program,
the PPE division of CERN,
the Swiss National Fund,
the INTAS under Contract INTAS-97-0158,
ORISE,
Grant-in-Aid for Scientific Research
(Specially Promoted Research \& International Scientific Research)
of the Ministry of Education, Science and Culture,
the University of Tsukuba Special Research Projects, and
the JSPS Research Fellowships for Young Scientists.
ORNL is managed by UT-Battelle, LLC, for the U.S. DOE
under contract DE-AC05-00OR22725.
The MIT group has been supported by the U.S. Dept. of Energy under the
cooperative agreement DE-FC02- 94ER40818.


\end{document}